\documentclass[pdflatex,sn-mathphys-num]{sn-jnl}

\usepackage{graphicx}%
\usepackage{multirow}%
\usepackage{amsmath,amssymb,amsfonts}%
\usepackage{amsthm}%
\usepackage{mathrsfs}%
\usepackage[title]{appendix}%
\usepackage{xcolor}%
\usepackage{textcomp}%
\usepackage{manyfoot}%
\usepackage{booktabs}%
\usepackage{algorithm}%
\usepackage{algorithmicx}%
\usepackage{algpseudocode}%
\usepackage{listings}%
\usepackage{lineno}

\theoremstyle{thmstyleone}%
\theoremstyle{thmstyletwo}%

\theoremstyle{thmstylethree}%

\begin{document}

\title[Phase Separation Bursting and Symmetry Breaking inside an Evaporating Droplet; Formation of a Flower-like Pattern]{Phase Separation Bursting and Symmetry Breaking inside an Evaporating Droplet; Formation of a Flower-like Pattern}


\author*[1,2]{\fnm{Vahid} \sur{Nasirimarekani}}\email{vahid.nasirimarekani@ds.mpg.de}

\affil*[1]{\orgdiv{Max Planck Institute for Dynamics and Self-Organization,}
\normalsize{Am Fassberg 17, 37077, Göttingen, Germany}}

\affil[2]{\orgdiv{Laboratory of Fluid Physics and Biocomplexity},
\normalsize{Am Fassberg 17, 37077, Göttingen, Germany}}


\abstract{Pattern formation inside a liquid phase is a phenomenon involved in many different aspects of life on our planet. The droplet form of a liquid that evaporates can reveal patterns that depend on the chemistry of the droplet and the physical parameters involved. We observed a flower-like deposition pattern of micrometer-sized particles as a result of the evaporation of a droplet containing salt and a non-ionic polymer. We show experimentally that the phase separation of the polymer due to the salting-out effect causes a strong entropic flow, which manifests as vortices. The flow is called “phase separation bursting flow”, which leads to the axial symmetry breaking and the formation of a radially aligned flower-like pattern. We foresee that understanding the observed flow can provide insights into the fluid physics aspects of phase separation and may have implications for technical applications.}




\maketitle
\newpage
Pattern formation in the liquid phase is a phenomenon characterized by the spontaneous emergence of ordered structures within a fluid system. This occurs through a variety of mechanisms, such as diffusion, convection, and surface tension effects, which drive the self-organization of particles, droplets, or other components within the liquid. These patterns can manifest as intricate assemblies observed in natural processes and experimental setups~\cite{cross1993pattern,gollub1999pattern}. The study of these patterns reveals underlying principles of non-equilibrium thermodynamics and fluid dynamics, providing insights into natural phenomena ranging from the growth of biological tissues to the dynamics of climate systems~\cite{landge2020pattern,boeynaems2018protein,koch1994biological,rietkerk2008regular}. 
\\
Pattern formation through phase separation is a fundamental topic in various scientific fields, including physics, chemistry, biology, and materials science. The concept revolves around understanding how distinct spatial structures emerge in systems such as a liquid phase that were initially homogeneous. Phase separation particularly was seen in biological systems, which results in pattern formation of organisms inside the liquid phase~\cite{liu2013phase}. Formation of membranless organelles is one of the examples of biomolecular patterning in microscopic scale due to liquid-liquid phase separation~\cite{boeynaems2018protein}. 
\\
When a homogeneous mixture of soluble components inside a liquid phase undergo non-equilibrium conditions, the components are entropically pushed to new spatial ordering in order to reach a new equilibrium stage. This process drives the system to a new state, which, if it continuously experiences non-equilibrium conditions, will continuously reshape its components accordingly~\cite{fang2019nonequilibrium}. There are many physical conditions that can put a liquid into a non-equilibrium dynamics, including the evaporation that any volatile liquid around us is subjected to evaporation~\cite{zudin2019non}. 
\\
A volatile liquid exposed to air with a relative humidity below the saturation point will evaporate, albeit very slowly. Evaporation influences internal dynamics of a liquid and potentially the pattern formation in the liquid phase by driving assembly of the particles within the liquid phase~\cite{heinen2019evaporation}. Moreover, as the surface area to volume ratio of the liquid increases (i.e. as the liquid volume decreases), the evaporation ratio and its effects on the liquid become dominant~\cite{gore2024symmetry}.
\\
When a small volume of a liquid evaporates in the form of a sessile droplet, the resulting entropic changes will cause fluid flows within the droplet, such as capillary flows~\cite{deegan1997capillary,yunker2011suppression} and Marangoni flows~\cite{weber1855mikroskopische,scriven1960marangoni, xu2007marangoni,hu2005analysis}. Marangoni effect arise from surface tension gradients caused by temperature or concentration differences~\cite{kim2022multiple}. These internal flows transport particles towards the droplet's edge or specific regions inside the droplet~\cite{deegan1997capillary}. Furthermore, increasing the complexity of the droplet by adding more components affects the dynamics of the droplet as it leads to changes in the final patterning of the droplet components~\cite{brutin2015droplet}.
\\
In this work, we conducted experiments around the question of possible pattern formation due to Marangoni vortex inside a evaporating droplet containing a non-volatile polymer and salt mixture. More specifically, how the interaction of the salt (another non-volatile component) with the polymer phase influences the final pattern formation inside the droplet, which exhibits a Marangoni flow. The schematics of the evaporating droplet is shown in Figure~\ref{fig:fig1}.a,b. 
\\
The experiments led to observing a regular pattern of the deposited particles on a glass substrate. A set of complementary experiments were conducted to disentangle the mechanism of flower-like pattern formation. By linking the dynamics of the particles inside the droplet with the chemical composition of the droplet, we introduce the concept of "phase separation bursting" as a mechanism which results in formation of vortices and patterning of the particles as a flower-like pattern. Considering that phase separation is one of the essential physical processes in nature, we foresee that the phase separation bursting flow could potentially lead us to understand the role of fluid physics associated to phase separation in naturally occurring patterns.
\begin{figure*}[h!]
    \centering
    \includegraphics[width=0.95\textwidth]{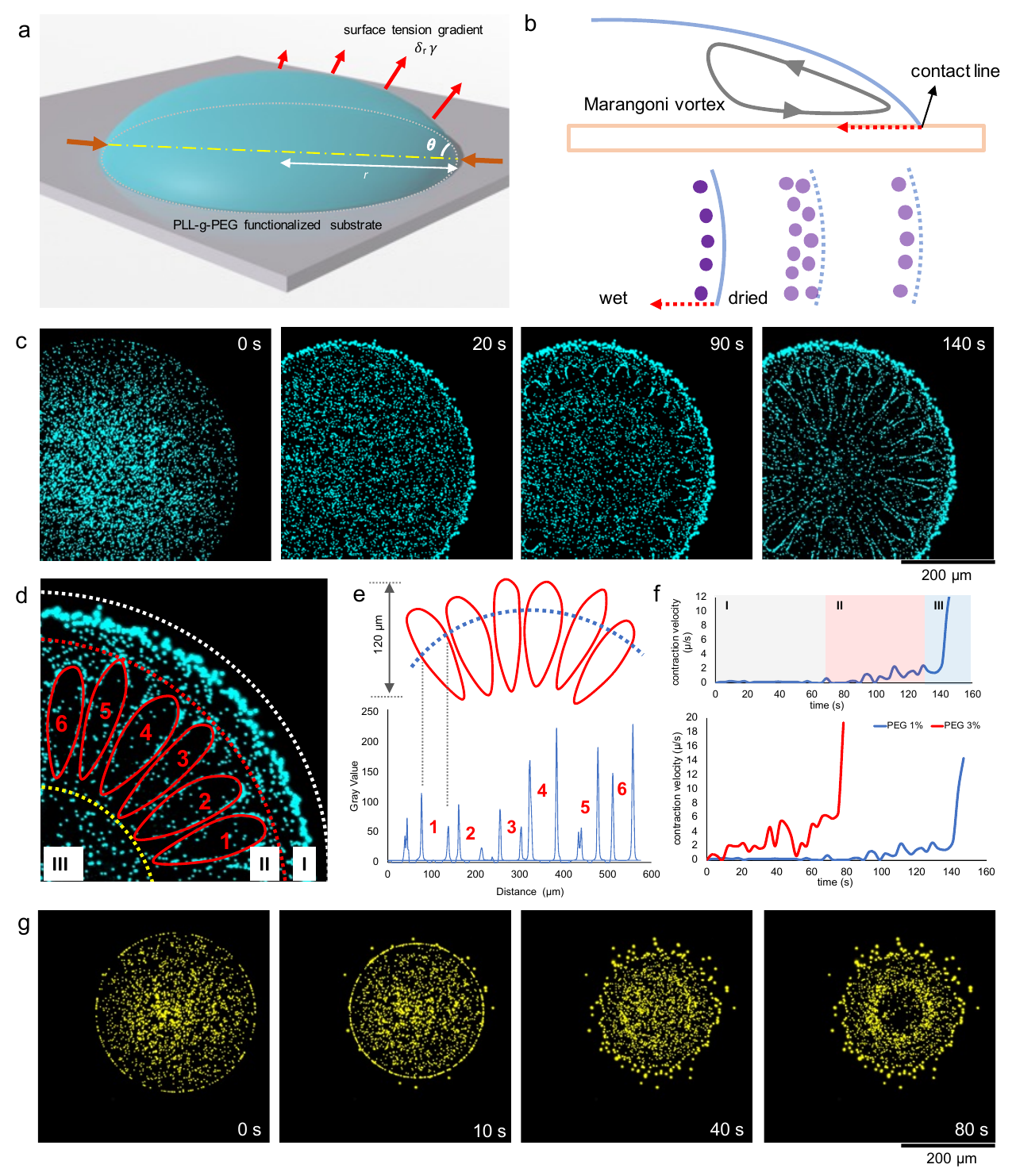}
    \caption{Symmetry breaking inside an evaporating droplet and deposition of spherical  beads as a flower-like pattern on a polymer-brushed coated glass substrate. a) 3D schematics of the evaporating sessile droplet on the substrate, highlighting the presence of surface tension gradient along the droplet. b) 2D schematics of the typical Marangoni vortex due to presence of polyethylene glycol(PEG) inside the droplet functioning as surfactant. The pattern formation is visualized by deposition pattern of the particles as droplet shrinks towards the droplet's center. c) Time-lapse images of the droplet containing 1\% w/w PEG in M2B buffer, resulting a flower-like deposition of the particles in the fully dried stage. d) An inset view depicting the flower-like pattern with highlighting the three different regions of the resulted pattern and the flower-like petals. e) Gray value plot of the flower pattern, indicating the wavelength of the petals, in the range of 30-70 $\mu m$. f) Contraction velocity of the droplet interface  as a function of the initial PEG concentration, 1\% w/w (upper) and comparison of 1\% with 3\% w/w (lower). g) Pattern formation in the droplet containing initial concentration of 3\% w/w PEG in M2B buffer.}
    \label{fig:fig1}
\end{figure*}

\section*{Results and Discussion}
A series of experiments were carried out in which droplets containing non-ionic polymer, polyethylene glycol (PEG), dissolved in salty buffer (M2B) and mixed with 2 $\mu m$ fluorescent beads (tracer particles) were subjected to drying on a functionalized glass substrate 
(see Methods). The droplets were deposited on a glass substrate with polymer brush coating of poly(L-lysine) backbone and poly(ethylene glycol) side-chains (PLL-g-PEG), in order to block any particle and polymer interaction with the substrate. The coating results in smooth contraction rather than pinning and de-pinning dynamics of the droplet throughout the course of evaporation. The M2B buffer was prepared by mixing monovalent and polyvalent salts to achieve a pH value around 6.9. The dynamics of the tracer particles over the course of drying and the final patterns were visualized by an inverted epi-fluorescent microscope. The main parameters considered relevant in the final pattern formation were: (i) initial PEG concentration (1\% and 3\% w/w concentrations), (ii) presence of salt, and (iii) PLL-g-PEG surface coating.  

\subsection*{Particle deposition as a flower-like pattern is a function of initial polymer concentration}
We have observed a regular and radially aligned polar ellipse shapes (hereinafter referred as flower-like pattern) formed by deposition of the tracer particles on the substrate, as the result of drying of a droplet containing 1\% w/w PEG in M2B buffer (Figure~\ref{fig:fig1}.c,d, Movie S1). The particle deposition can be divided into three different regions, namely: (i) the outer ring-like pattern with a random deposition of particles, (ii) the intermediate region wherein the flower-like pattern forms and (iii) the central region which shows an-isotropic deposition (Figure~\ref{fig:fig1}.c). The separation of the regions is defined by formation of the flower-like petals which makes an intermediate region in between the outer and inner part of the deposited patterns (shown in red in Figure~\ref{fig:fig1}.d).
\\
Transition of the symmetric ring-like deposition to the radial flower-like pattern shows occurrence a axial symmetry breaking mechanism. The petals form radial domains with a minimum gap of a few microns between them and a degree of variation in size and shape (Figure~\ref{fig:fig1}.d,e). The gray value measurements, which show the intensity of the detected light as a function of the particle concentration, show the wavelength of the petals in the range of 30-70 $\mu m$. The length of the petals have an average value of 120 $\mu m$, for a droplet of around 300 $\mu m$ of radius. In addition, repeating the experiments with a concentration of 1\% exhibits that the symmetry breaking mechanism is altered to some extent; therefore, the flower pattern in the final sample may have some deviations, and the petal formation might be incomplete (Figure S1).
\\
The observed flower-like patterns form at the interface of the evaporating droplet. The deposition of the particles occurs on the substrate at the interface of the evaporating droplet, while the wet part of the droplet contracts towards its center (Figure~\ref{fig:fig1}.b). This raises the question of the effect of the contraction velocity of the interface on the final pattern. The contraction velocity of the contact line of the droplet, measured with respect to the initial contact line of the droplet, exhibits different velocity profiles in each region within the droplet (Figure~\ref{fig:fig1}.f, upper). It shows relatively linear contraction for the ring-like region (I), an increasing trend with considerable oscillations in the flower-like pattern region (II), and relatively fast contraction in the central region (III). 
\\
Considering that the strength of the Marangoni flow can alter the contraction dynamics of the droplet~\cite{thokchom2019dynamical}, we compared the contraction velocity profiles for droplets with 1\% and 3\% w/w PEG concentrations (Figure~\ref{fig:fig1}.f, lower). The 3\% droplet showed stronger contraction and more pronounced oscillations in the II region. Nevertheless, similar trends and velocity profile is seen as the 1\% droplet case. Surprisingly, the particle pattern in 3\% PEG droplet reveals no symmetry breaking and flower-like pattern formation (Figure~\ref{fig:fig1}.g). The 3\% PEG droplet shows stronger Marangoni vortices at the contact line than the 1\% PEG droplet (Movie S2). The increase in the initial concentration and consequently the strength of the Marangoni vortex inside the droplet leads to a different pattern formation of the particles. Nevertheless, the strength of the Marangoni vortex does not explain the mechanism behind the formation of the flower-like pattern~\cite{parsa2018mechanisms}.
\\
The flower-like pattern formation and symmetry breaking inside a droplet is reported in literature~\cite{wodlei2018marangoni,yamamoto2015evolution}, but as Marangoni spreading of a liquid droplet on top of liquid phase substrate. Marangoni spreading~\cite{chan2024marangoni,baumgartner2022marangoni,ma2023experiments,ma2020fingering} explains the patterning of the small droplets in which the substrate liquid phase repels the droplets in de-wetting phase due to the surface tension. However, in our experiments, the droplet shrinks on a solid substrate which has inward motion towards the center of the droplet. In similar work reported in the literature, the Marangoni vortex inside an evaporating sessile droplet shows only the formation of bands in radial symmetry at the contact line~\cite{thokchom2019dynamical}. In addition, the reported work shows a continuous pinning and de-pinning dynamic that eventually leads to the formation of multiple coffee rings~\cite{thayyil2022evaporation,bi2012unconventional}.
\\
It is important to note that our droplet contains a salt mixture, which adds additional complexity to the droplet. However, the intriguing aspect is that regardless of the complexity of the system, a flower-like pattern can be seen that resembles the Marangoni spreading effect. To understand the physics behind the emergence of the flower-like pattern, we have followed a top-down approach to elucidate the mechanism of the flower-like patterning of the particles. In the following sections, by reducing the complexity of the droplet, the effect of each component on the dynamics of the droplet and the final pattern formation is investigated and discussed.  
\begin{figure*}[tb!]
    \centering
    \includegraphics[width=\textwidth]{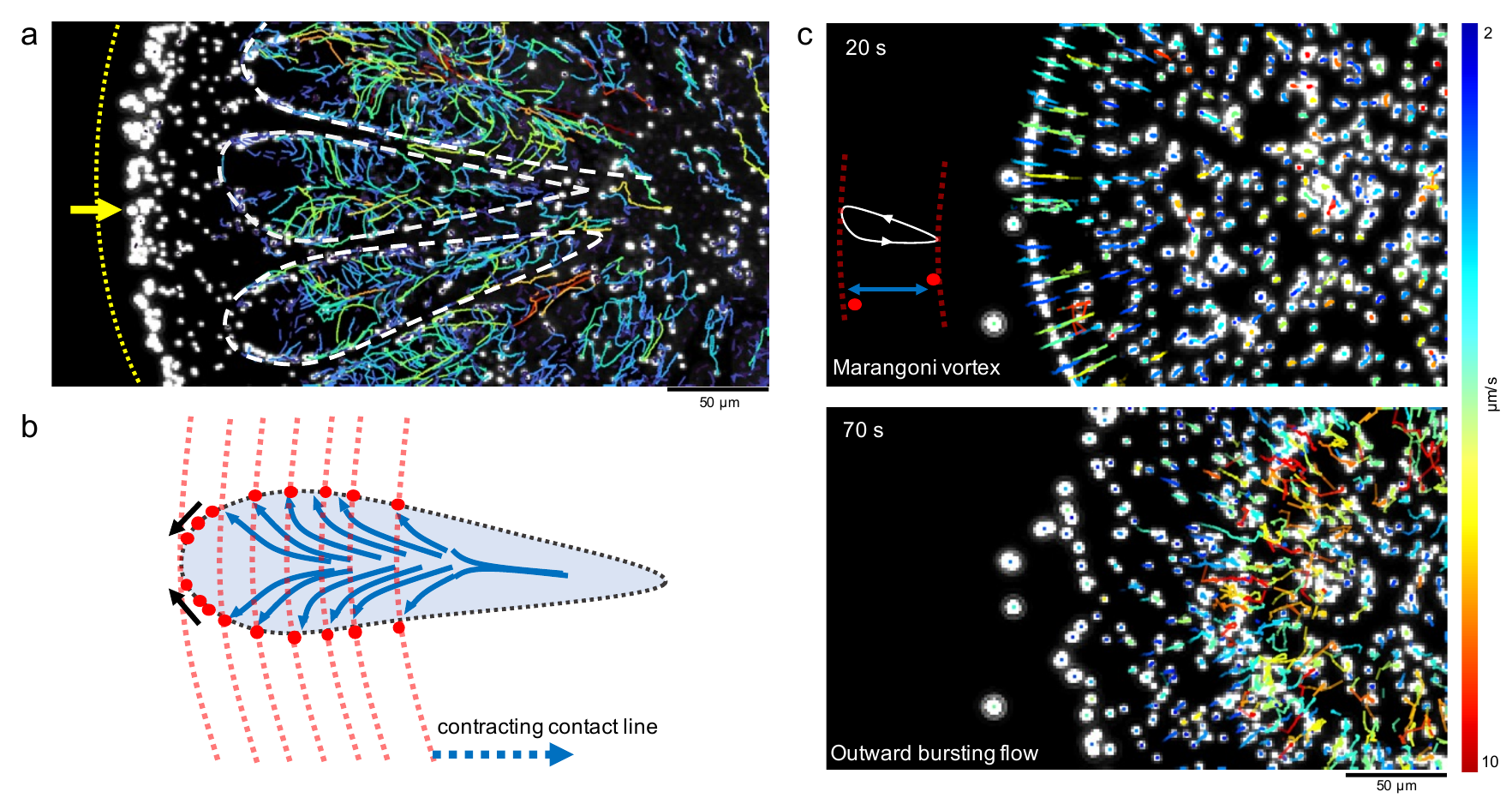}
    \caption{Dynamics of particle deposition and formation of the flower-like pattern inside the evaporating sessile droplet. a) An inset of the droplet containing 1\% w/w PEG with plotted trajectories of the particles over time in the flower-like pattern formation region. b) Schematics of the mechanisms of the flower petal formation based on the particle trajectories, red dash-lines represent the contracting contact line moving towards the droplet center. The blue arrows show the vortices, which propel the particle towards the droplet's interface. c) An inset of the droplet containing 3\% w/w PEG, It emphasizes the presence of the Marangoni vortex in the early stages of evaporation and the appearance of the outward bursting flow at a later stage.}
    \label{fig:fig2}
\end{figure*}
\subsection*{The axial symmetry breaks and flower-like patterns form due to vortices inside the evaporating droplet}
Since the flower-like pattern is the result of the deposition of spherical beads during the drying process, we performed particle tracking to observe the dynamics of the axial symmetry breaking.
\\
The particle trajectories shows that the flower-like pattern forms due to on-plane vortices resembling a fountain flow (Figure~\ref{fig:fig2}.a,b). The paired vortices propel the particles in two opposite directions, analogous to a fountain. At the same time, the contact line moves towards the center of the droplet due to the continuous evaporation and contraction of the droplet. As a result, the particles pin to the substrate at the contact line, which leads to the formation of the petal-like shape (Movie S3). In other words, the particles in two opposing vortex flows reach the contact line of the droplet, which moves towards the center of the droplet, causing the particles to be deposited (Figure~\ref{fig:fig2}.b, the course of the contact line is shown as dashed lines).  
\\
Increasing the initial PEG concentration (3\% PEG droplet) shows there are two main flows inside the evaporating droplet: (i) a typical Marangoni vortex circulation in the early stages of evaporation, and (ii) an outward bursting flow of particles in the later stages of evaporation (Figure\ref{fig:fig2}.c). This indicates that these two different flows contribute to the patterning of the particles one after the other. The Marangoni flow reaches a maximum and then appears to stop. Then the bursting flow, which is stronger than the Marangoni flow, emerges in the central part of the droplet. The Marangoni vortex is normally expected in a PEG-containing evaporating droplet and is a function of the initial PEG concentration~\cite{seo2017altering}. In contrast, the outward flow that breaks axial symmetry is not expected to be caused by the surface tension gradient induced by PEG.
\\
We hypothesize that in 1\% PEG droplet, the strength of the Marangoni flow and consequently the contraction ratio favors flower-like deposition. It provides sufficient spatial freedom for uniform deposition of the particles, which are pushed by the outward repulsive flow. The Marangoni strength as a function of the initial polymer concentration determines the contraction ratio of the droplet, with a higher contraction being achieved by increasing the PEG concentration (Figure~\ref{fig:fig1}.f). The higher and faster contraction rate of the 3\% PEG droplet concentrates the particles in the center, which limits the spatial freedom for the bursting flow to result a regular pattern. We therefore assume that these two flows interact and determine the final patterning of the particles.
\\
The observation of vortices within an evaporating sessile droplet is reported in the literature as solutal Marangoni flow, which occurs at high salt concentrations and with inhomogeneous evaporation of the droplet~\cite{bennacer2014vortices,kim2017solutal,pyeon2022self}. The underlying physics involves surface tension gradients caused by the non-uniform distribution of solutes during evaporation. The difference in surface tension across the droplet’s surface causes fluid to flow from regions of low surface tension to regions of high surface tension~\cite{diddens2024non}. The driven flow induces internal circulation within the droplet, drawn from the edges towards the center on the surface and creating a convective loop~\cite{park2020control,ryu2021analysis}. These convective loops can manifest as vortices inside the droplet. In some cases, multiple vortices can form depending on the droplet's size, solute concentration, and evaporation conditions. Considering the presence of salt in our droplet together with PEG, The emergence of these two different flows was anticipated. However, both components are non-volatile, which means that the interaction between these two components and their deposition in the same droplet at later stages of evaporation must be understood. In other words, the chemistry of the droplet will determine how these components occupy the space inside the droplet. Answering this question is therefore crucial and could explain the shift in flow from a Marangoni vortex to an outward flow as evaporation progresses.
\subsection*{The phase separation caused by the salting-out effect leads to a “phase separation bursting” flow and the formation of axial vortices}
\begin{figure*}[tb!]
    \centering
    \includegraphics[width=\textwidth]{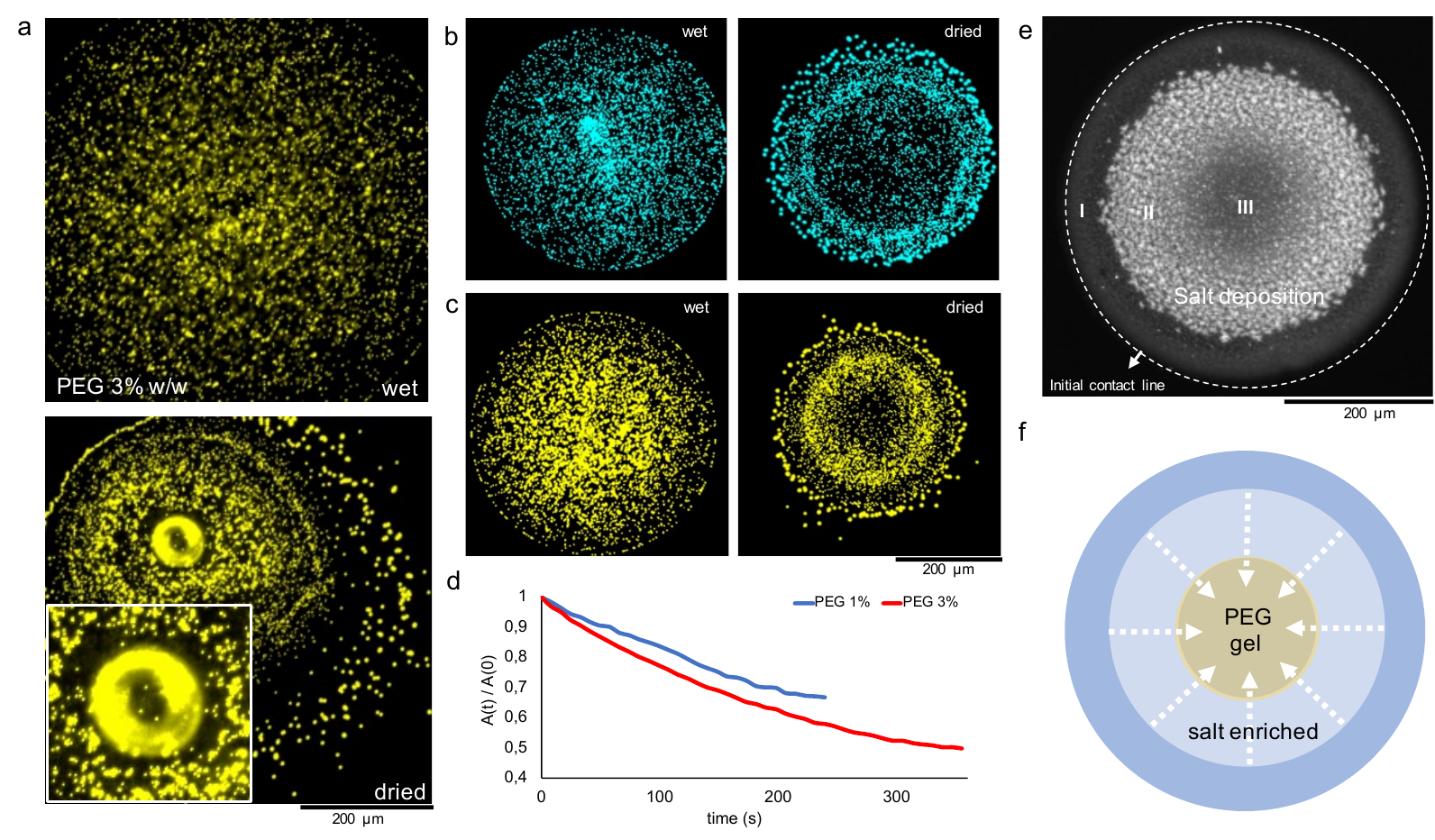} 
    \caption{The role of salt in the formation of bursting flow and pattern formation within the evaporating sessile droplet. a) Formation of PEG gel in the center of the droplet at later stages of evaporation, the yellow  donut shape shows the PEG gel bump. b,c) Microscopic images of the evaporating droplet containing 1\% and 3\% w/w PEG dried on Pll-g-PEG functionalized glass. d) Contraction ratio of PEG droplets with 1\% and 3\% w/w, in deionized water as buffer(A(t)/A(0) measures the surface area ratio of the droplet in relation to the initial surface area). e) Microscopic image of salt deposition as a result of evaporation of the salt-containing buffer, M2B. f) 2D schematic showing the spatial deposition of PEG and salt crystals in the fully evaporated droplet, with PEG accumulating in the center and salt as a ring around the PEG.}
    \label{fig:fig3}
\end{figure*}

Drying of the droplet with initial polymer concentration of 3\% leads to observe a macroscopic phase transition of the polymer as a gel phase in the later stages of evaporation (Figure~\ref{fig:fig3}.a). A clear circular bump shape of the PEG gel was observed in 3\% PEG droplet which was not seen in 1\% PEG droplet. As the droplet evaporates, the PEG concentration increases over time. In early stages of evaporation, the PEG is soluble in the buffer and due to evaporation reaches the concentration favorable for emergence of Marangoni vortex. Upon further increasing the concentration over certain threshold (around 20-30\%), it undergoes a phase transition known as sol-gel transition~\cite{li1994sol,baldwin2014effects}, which brings the PEG to the center of the droplet and finally forming the compact gel deposition. The phase transition suggests that prior to sol-gel transition, the PEG undergoes a phase separation which results in PEG rich regions inside the droplet~\cite{asenjo2012aqueous}.
\\
The observed phase transition (accumulation of PEG gel in the center of the droplet) suggests the role of salt on driving the mixture to the sol-gel transition. Phase separation in polymer solutions due to the addition of salts is known as salting-out effect. When salts are added to a polymer solution, the salt ions preferentially interact with the water molecules, effectively "stealing" the hydration shell from the polymer chains~\cite{grover2005critical,arakawa1984mechanism}. Continuous evaporation results in increment of the salt/water ratio inside the droplet, therefore, most of the water molecules get involved in hydrating the salt ions, reducing the number of water molecules available to the polymer. The dehydration of PEG chains due to the preferential hydration of salt ions reduces the polymer's solubility.
This leads to an increase in PEG-PEG interactions, resulting in phase separation~\cite{hey2005salting}. 
\\
Excluding salt from the buffer was thought to give a better insight for the flower-like pattern formation. In this regard, we have conducted experiments with PEG in deionized water. The results show absence of the vortices and no regular pattern formation in both 1\% and 3\% PEG droplets (Figure~\ref{fig:fig3}.b,c). However, the contraction ratio (A(t)/A(0)) of the droplet is similar to the previous experiments, showing a higher contraction ratio for 3\% PEG in comparison to 1\% PEG droplet (Figure~\ref{fig:fig3}.d). Moreover, the Marangoni flow was more visible in the case of 3\% PEG droplet which resulted in the ring shape deposition of the particles in the center of the droplet (Figure~\ref{fig:fig3}.c, dried). The particle tracking show no visible vortices in both concentrations in the absence of salt (Movie S4, S5). In addition, the dynamics of particles and final pattern formation in the absence of salt is very similar to reported works in literature~\cite{seo2017altering,thokchom2019dynamical}; the strength of Marangoni flow increases over the course of drying and particles are deposited in the center of the droplet. These observations indicate that the presence of salt caused the vortices to be formed. 
\\
On the other side, excluding PEG from the droplet shows that the salt crystals form in the region II, where the flower-like pattern was previously formed (Figure \ref{fig:fig3}.e). Moreover, drying the salty buffer with tracer particles does not show any vortices (Figure S2). 
By combining the two observations, the gel formation in the III region and the salt deposition in the II region, we conclude that there is a phase separation between the salt crystals and the polymer gel (Figure \ref{fig:fig3}.f). Therefore, salt crystallization in the presence of PEG leads to an outward repulsive flow and the formation of vortices. We assume that this means that the salt competes for space in region II and pushes the PEG into the center of the droplet (Figure \ref{fig:fig3}.f). 
\\
Another indication of the salting-out effect is the repulsion of PEG by the salt molecules near the contact line, where PEG accumulates to reduce the surface tension of the droplet. It has been reported that the addition of salt reduces the surface tension~\cite{wu1996interfacial}, as PEG is further deposited at the interface in the presence of salt~\cite{qazi2017influence}. Our measurements confirm that the M2B buffer reduced the surface tension of 1\% PEG solution from 62.95 to 61.34 mN/m~\cite{nasirimarekani2023pattern}. Due to the repulsion between PEG and salt, they are eventually pushed away from each other before the phase transition to the gel phase.
\\
From a thermodynamic point of view, the observed vortices or outward repulsive flows are caused by the entropic effect~\cite{falahati2019thermodynamically,shrinivas2021phase,steiner1995entropy}. The entropic effect occurs when the mixed volume attempts to transition into a spatially separated phase. This transition induces entropic forces that cause the particles to be pushed out of the polymer phase, which undergoes phase separation. Therefore, we refer to this flow as “phase separation bursting” flow, which leads to long-range vortices in a confined volume of liquid, here in a sessile droplet.
\subsection*{Polymer brushes on the substrate facilitate phase separation and minimize particle-substrate interaction}
\begin{figure}[h]
    \centering
    \includegraphics[width=0.7\textwidth]{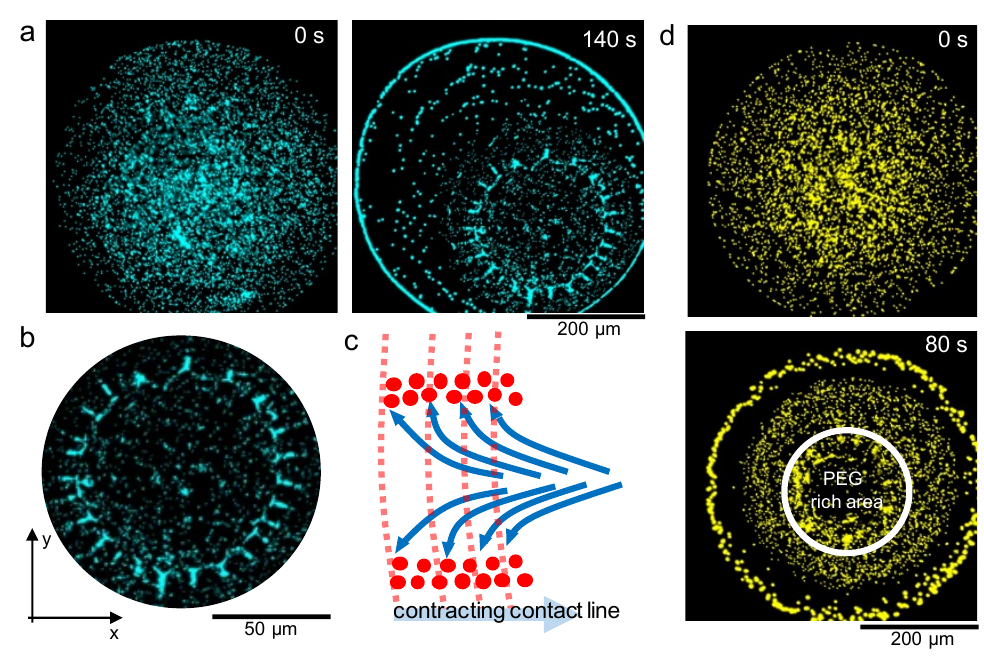}
    \caption{Pattern formation on glass substrate without polymer brushes. a) Deposition pattern of the particles inside the droplet of 1\% PEG. b) An inset view of the pattern formed in the center of the droplet, resembling a fan shape. c) Schematics of the bursting vortices which deposit the particles as the fan shape. d) Pattern formation and observing of the PEG accumulation in the 3\% PEG droplet}
    \label{fig:fig4}
\end{figure}

In the previous experiments, the glass cover substrate was functionalized  with PLL-g-PEG brushes, which limits the interaction of the polymer monomers and the particles with the substrate and causes a continuous contraction of the droplet. This eliminates the pinning and de-pinning dynamics of the interface during evaporation (Figure~\ref{fig:fig3}.d, seen as a continuous and gradual contraction). These properties of PEG brushes raise the question of their role in flow dynamics and final pattern formation. Therefore, the same experiment as in Figure~\ref{fig:fig1} (PEG concentration of 1\% and 3\% in M2B buffer) was performed on a clean glass substrate.
\\
The results show a polar clustering of particles in a circular pattern in the center of the droplet, analogous to a fan shape in the case of a 1\% PEG droplet (Figure~\ref{fig:fig4}.a,b). Tracing the particle trajectories over time shows the presence of both the previously observed Marangoni vortex and the repulsive flow in the center of the droplet (Figure S3).  Particle tracks show that the fan-shaped pattern in the center of the droplet is the result of the repulsive vortices, similar to the flower-like pattern formation (Figure~\ref{fig:fig4}.c and Figure S3). 
The particles accumulate in the form of rod-shaped fingers in the phase separation step (Figure~\ref{fig:fig4}.c). A typical coffee ring is also visible at the initial contact line of the droplet. The time-lapse videos of the experiment show that the particles adhere to the glass substrate as the droplet contracts during the drying process.
\\
On the one hand, the difference in structuring on the clean glass substrate is due to the interaction of the particles with the substrate, which leads to additional friction for their movement and deposition. The friction limits the repulsion flow and the deposition of the particles over a greater distance, resulting in a relatively compact arrangement of the particles~\cite{zeuthen2019nanoparticle}. 
On the other hand, the polymer brushes also prevent the loss of PEG monomers binding on the substarte and accumulation on the glass substrate during the drying process, due to the antifouling property of the polymer brushes~\cite{higaki2016anti}. In the case of functionalized glass, there are repulsive forces between the free PEG monomers in the droplet and the PLL-g-PEG brushes on the substrate. This repulsion prevents the PEG molecules from binding to the substrate. Therefore, it facilitates the phase separation of PEG in later stages of evaporation. Comparing the 3\% PEG with previous cases, we see that it is a PEG-rich deposit rather than a compact gel in the center of the droplet.  We conclude that the PLL-g-PEG brushes form a thin lubricant layer on the substrate~\cite{kreer2016polymer}, which restricts the interaction between particles and substrate and facilitates the formation of PEG gels in the center of the droplet. Although the main flows, the Marangoni and phase separation flows, are not influenced by the surface properties, the final pattern is still affected, highlighting the role of the substrate in the phase transition of the polymer and in the final patterning of the particles.
\\
It is important to note that PLL-g-PEG brushes also lead to a decrease in the contact angle, resulting in a smaller depth of the droplet. Contact angle measurements show the initial contact angles of $16\pm0.5$ and $10\pm1.4$ for clean glass and functionalized glass, respectively (in the case of 1\% PEG droplets). The polymer brushes improves the wetting properties of the substrate, which indicates the lubricity of the surface due to the polymer brushes lowering the contact angle~\cite{giasson2014aqueous}. A higher contact angle affects the dynamics in the early stages of drying, as more particles can accumulate at the contact line, leading to the formation of clear coffee rings (Figure~\ref{fig:fig4}.a).
\section*{Conclusion}
This article discusses the presence of a bursting flow in an evaporation droplet, referred to here as “phase separation bursting” flow. The flow is caused by phase separation in an evaporation droplet in which the polymeric monomers begin to form a gel phase. The phase separation caused by the salting-out effect leads to the formation of microscopically visible vortices. The force for the flow is generated by abrupt entropic changes as a result of the phase separation. The vortices then break the axial symmetry and push the particles, depositing them in a random orientation or a flower-like pattern, depending on the initial concentration of the polymer in the droplet.  These results indicate the potential of phase separation in generating flows over large distances in small volumes. These findings could provide important insights into the formation of similar patterns, as phase separation is an intrinsic aspect of nature. As a further step towards a better understanding of phase separation flow, we propose to investigate the role of salt in more detail, as different ionic strengths could influence the extent of the phase separation and the entropic forces.
\newpage
\section*{Material and Methods}
\subsection*{Functionalized glass surface preparation}
 Microscope cover-slips were cleaned by washing with 100\% ethanol and rinsing in deionized water. They were further sonicated in acetone for 30 min and incubated in ethanol for 10 min at room temperature. This was followed by incubation in a 2\% Hellmanex III solution (Hellma Analytics) for 2 h, extensive washing in deionized water, and drying with a filtered airflow. The cleaned cover-slips were immediately activated in
oxygen plasma (FEMTO, Diener Electronics, Germany) for 30 s at 0.5 mbar. 0.1 mg/ml Poly(L-lysine)-graft-poly(ethylene glycol) (PLL-g-PEG) (SuSoS AG, Switzerland) in 10 mM HEPES (pH 7.4, at room temperature) was poured onto the glass cover slip and incubated for 1 h, finally pressurized dry air stream was used to remove the liquid and dry the glass cover-slips.
\subsection*{Droplet composition}
The PEG droplet contained M2B buffer (80 mM PIPES, adjusted to pH = 6.9 with KOH, 1 mM EGTA, 2 mM MgCl$_{2}$) were mixed with fluorescent beads (2 $\mu m$ in diameter, red fluorescent 580/605 were purchased from Invitrogen™). M2B buffer was used as the basic buffer to solve 20 kDa PEG polymer. Moreover, the PEG solution was also prepared in de-ionized water for experiments without salt. 
\subsection*{Image Acquisition}
Image acquisition was performed using an inverted fluorescence microscope Olympus IX-71 with a 4$\times$ objective (Olympus, Japan) or 60$\times$ oil-immersion objective (Olympus, Japan), depending on the experimental setup. For excitation, a Lumen 200 metal arc lamp (Prior Scientific Instruments, U.S.A.) was applied. The images were recorded with a CCD camera (CoolSnap HQ2, Photometrics). The frames were acquired with a variable rate according to the experiment with an exposure time of 300 ms for a variable time according to the experiment.
\subsection*{Image Analysis}
Imagej software was used to process the acquired images sequences. The particle tracking was conducted with using Trackmate plugin in Imagej with customized values to achieve accurate tracking trajectories. The data of particle tracking was processed with a self-developed matlab code to plot the flow map of the particle trajectory over the course of drying. The speed profiles were calculated and plotted on the basis of the tracking data.  
\section*{Acknowledgement}
The author acknowledges the critical and valuable comments of Prof. Bodenschatz. Additionally, I would like to thank Katharina Gunkel and Lasse Lehmann for their assistance in preparing the material. This project was supported by funding of Max Planck Society and the Klaus Tschira Boost Fund, a joint initiative of the German Scholars Organization and the Klaus Tschira Foundation.
\section*{Author Contributions}
As a single author, the entire work, from the conception and execution of the experiments to writing the manuscript, was carried out by the author himself. 
\section*{Competing Interests}
The authors declare no competing interests.
\bibliography{sn-bibliography}


\begin{thebibliography}{53}
\ifx \bisbn   \undefined \def \bisbn  #1{ISBN #1}\fi
\ifx \binits  \undefined \def \binits#1{#1}\fi
\ifx \bauthor  \undefined \def \bauthor#1{#1}\fi
\ifx \batitle  \undefined \def \batitle#1{#1}\fi
\ifx \bjtitle  \undefined \def \bjtitle#1{#1}\fi
\ifx \bvolume  \undefined \def \bvolume#1{\textbf{#1}}\fi
\ifx \byear  \undefined \def \byear#1{#1}\fi
\ifx \bissue  \undefined \def \bissue#1{#1}\fi
\ifx \bfpage  \undefined \def \bfpage#1{#1}\fi
\ifx \blpage  \undefined \def \blpage #1{#1}\fi
\ifx \burl  \undefined \def \burl#1{\textsf{#1}}\fi
\ifx \doiurl  \undefined \def \doiurl#1{\url{https://doi.org/#1}}\fi
\ifx \betal  \undefined \def \betal{\textit{et al.}}\fi
\ifx \binstitute  \undefined \def \binstitute#1{#1}\fi
\ifx \binstitutionaled  \undefined \def \binstitutionaled#1{#1}\fi
\ifx \bctitle  \undefined \def \bctitle#1{#1}\fi
\ifx \beditor  \undefined \def \beditor#1{#1}\fi
\ifx \bpublisher  \undefined \def \bpublisher#1{#1}\fi
\ifx \bbtitle  \undefined \def \bbtitle#1{#1}\fi
\ifx \bedition  \undefined \def \bedition#1{#1}\fi
\ifx \bseriesno  \undefined \def \bseriesno#1{#1}\fi
\ifx \blocation  \undefined \def \blocation#1{#1}\fi
\ifx \bsertitle  \undefined \def \bsertitle#1{#1}\fi
\ifx \bsnm \undefined \def \bsnm#1{#1}\fi
\ifx \bsuffix \undefined \def \bsuffix#1{#1}\fi
\ifx \bparticle \undefined \def \bparticle#1{#1}\fi
\ifx \barticle \undefined \def \barticle#1{#1}\fi
\bibcommenthead
\ifx \bconfdate \undefined \def \bconfdate #1{#1}\fi
\ifx \botherref \undefined \def \botherref #1{#1}\fi
\ifx \url \undefined \def \url#1{\textsf{#1}}\fi
\ifx \bchapter \undefined \def \bchapter#1{#1}\fi
\ifx \bbook \undefined \def \bbook#1{#1}\fi
\ifx \bcomment \undefined \def \bcomment#1{#1}\fi
\ifx \oauthor \undefined \def \oauthor#1{#1}\fi
\ifx \citeauthoryear \undefined \def \citeauthoryear#1{#1}\fi
\ifx \endbibitem  \undefined \def \endbibitem {}\fi
\ifx \bconflocation  \undefined \def \bconflocation#1{#1}\fi
\ifx \arxivurl  \undefined \def \arxivurl#1{\textsf{#1}}\fi
\csname PreBibitemsHook\endcsname

\bibitem[\protect\citeauthoryear{Cross and Hohenberg}{1993}]{cross1993pattern}
\begin{barticle}
\bauthor{\bsnm{Cross}, \binits{M.C.}},
\bauthor{\bsnm{Hohenberg}, \binits{P.C.}}:
\batitle{Pattern formation outside of equilibrium}.
\bjtitle{Reviews of modern physics}
\bvolume{65}(\bissue{3}),
\bfpage{851}
(\byear{1993})
\end{barticle}
\endbibitem

\bibitem[\protect\citeauthoryear{Gollub and Langer}{1999}]{gollub1999pattern}
\begin{barticle}
\bauthor{\bsnm{Gollub}, \binits{J.P.}},
\bauthor{\bsnm{Langer}, \binits{J.S.}}:
\batitle{Pattern formation in nonequilibrium physics}.
\bjtitle{Reviews of Modern Physics}
\bvolume{71}(\bissue{2}),
\bfpage{396}
(\byear{1999})
\end{barticle}
\endbibitem

\bibitem[\protect\citeauthoryear{Landge et~al.}{2020}]{landge2020pattern}
\begin{barticle}
\bauthor{\bsnm{Landge}, \binits{A.N.}},
\bauthor{\bsnm{Jordan}, \binits{B.M.}},
\bauthor{\bsnm{Diego}, \binits{X.}},
\bauthor{\bsnm{M{\"u}ller}, \binits{P.}}:
\batitle{Pattern formation mechanisms of self-organizing reaction-diffusion systems}.
\bjtitle{Developmental biology}
\bvolume{460}(\bissue{1}),
\bfpage{2}--\blpage{11}
(\byear{2020})
\end{barticle}
\endbibitem

\bibitem[\protect\citeauthoryear{Boeynaems et~al.}{2018}]{boeynaems2018protein}
\begin{barticle}
\bauthor{\bsnm{Boeynaems}, \binits{S.}},
\bauthor{\bsnm{Alberti}, \binits{S.}},
\bauthor{\bsnm{Fawzi}, \binits{N.L.}},
\bauthor{\bsnm{Mittag}, \binits{T.}},
\bauthor{\bsnm{Polymenidou}, \binits{M.}},
\bauthor{\bsnm{Rousseau}, \binits{F.}},
\bauthor{\bsnm{Schymkowitz}, \binits{J.}},
\bauthor{\bsnm{Shorter}, \binits{J.}},
\bauthor{\bsnm{Wolozin}, \binits{B.}},
\bauthor{\bsnm{Van Den~Bosch}, \binits{L.}}, \betal:
\batitle{Protein phase separation: a new phase in cell biology}.
\bjtitle{Trends in cell biology}
\bvolume{28}(\bissue{6}),
\bfpage{420}--\blpage{435}
(\byear{2018})
\end{barticle}
\endbibitem

\bibitem[\protect\citeauthoryear{Koch and Meinhardt}{1994}]{koch1994biological}
\begin{barticle}
\bauthor{\bsnm{Koch}, \binits{A.-J.}},
\bauthor{\bsnm{Meinhardt}, \binits{H.}}:
\batitle{Biological pattern formation: from basic mechanisms to complex structures}.
\bjtitle{Reviews of modern physics}
\bvolume{66}(\bissue{4}),
\bfpage{1481}
(\byear{1994})
\end{barticle}
\endbibitem

\bibitem[\protect\citeauthoryear{Rietkerk and Van~de Koppel}{2008}]{rietkerk2008regular}
\begin{barticle}
\bauthor{\bsnm{Rietkerk}, \binits{M.}},
\bauthor{\bsnm{Koppel}, \binits{J.}}:
\batitle{Regular pattern formation in real ecosystems}.
\bjtitle{Trends in ecology \& evolution}
\bvolume{23}(\bissue{3}),
\bfpage{169}--\blpage{175}
(\byear{2008})
\end{barticle}
\endbibitem

\bibitem[\protect\citeauthoryear{Liu et~al.}{2013}]{liu2013phase}
\begin{barticle}
\bauthor{\bsnm{Liu}, \binits{Q.-X.}},
\bauthor{\bsnm{Doelman}, \binits{A.}},
\bauthor{\bsnm{Rottsch{\"a}fer}, \binits{V.}},
\bauthor{\bsnm{Jager}, \binits{M.}},
\bauthor{\bsnm{Herman}, \binits{P.M.}},
\bauthor{\bsnm{Rietkerk}, \binits{M.}},
\bauthor{\bsnm{Koppel}, \binits{J.}}:
\batitle{Phase separation explains a new class of self-organized spatial patterns in ecological systems}.
\bjtitle{Proceedings of the National Academy of Sciences}
\bvolume{110}(\bissue{29}),
\bfpage{11905}--\blpage{11910}
(\byear{2013})
\end{barticle}
\endbibitem

\bibitem[\protect\citeauthoryear{Fang et~al.}{2019}]{fang2019nonequilibrium}
\begin{barticle}
\bauthor{\bsnm{Fang}, \binits{X.}},
\bauthor{\bsnm{Kruse}, \binits{K.}},
\bauthor{\bsnm{Lu}, \binits{T.}},
\bauthor{\bsnm{Wang}, \binits{J.}}:
\batitle{Nonequilibrium physics in biology}.
\bjtitle{Reviews of Modern Physics}
\bvolume{91}(\bissue{4}),
\bfpage{045004}
(\byear{2019})
\end{barticle}
\endbibitem

\bibitem[\protect\citeauthoryear{Zudin and Zudin}{2019}]{zudin2019non}
\begin{bbook}
\bauthor{\bsnm{Zudin}, \binits{Y.B.}},
\bauthor{\bsnm{Zudin}}:
\bbtitle{Non-equilibrium Evaporation and Condensation Processes}.
\bpublisher{Springer}, \blocation{???}
(\byear{2019})
\end{bbook}
\endbibitem

\bibitem[\protect\citeauthoryear{Heinen and Vrabec}{2019}]{heinen2019evaporation}
\begin{botherref}
\oauthor{\bsnm{Heinen}, \binits{M.}},
\oauthor{\bsnm{Vrabec}, \binits{J.}}:
Evaporation sampled by stationary molecular dynamics simulation.
The Journal of Chemical Physics
\textbf{151}(4)
(2019)
\end{botherref}
\endbibitem

\bibitem[\protect\citeauthoryear{Gore et~al.}{2024}]{gore2024symmetry}
\begin{botherref}
\oauthor{\bsnm{Gore}, \binits{S.}},
\oauthor{\bsnm{Paudyal}, \binits{B.}},
\oauthor{\bsnm{Ali}, \binits{M.}},
\oauthor{\bsnm{Masmoudi}, \binits{N.}},
\oauthor{\bsnm{Bae}, \binits{A.}},
\oauthor{\bsnm{Steinbock}, \binits{O.}},
\oauthor{\bsnm{Gholami}, \binits{A.}}:
Symmetry breaking in chemical systems: Engineering complexity through self-organization and marangoni flows.
arXiv preprint arXiv:2406.18006
(2024)
\end{botherref}
\endbibitem

\bibitem[\protect\citeauthoryear{Deegan et~al.}{1997}]{deegan1997capillary}
\begin{barticle}
\bauthor{\bsnm{Deegan}, \binits{R.D.}},
\bauthor{\bsnm{Bakajin}, \binits{O.}},
\bauthor{\bsnm{Dupont}, \binits{T.F.}},
\bauthor{\bsnm{Huber}, \binits{G.}},
\bauthor{\bsnm{Nagel}, \binits{S.R.}},
\bauthor{\bsnm{Witten}, \binits{T.A.}}:
\batitle{Capillary flow as the cause of ring stains from dried liquid drops}.
\bjtitle{Nature}
\bvolume{389}(\bissue{6653}),
\bfpage{827}--\blpage{829}
(\byear{1997})
\end{barticle}
\endbibitem

\bibitem[\protect\citeauthoryear{Yunker et~al.}{2011}]{yunker2011suppression}
\begin{barticle}
\bauthor{\bsnm{Yunker}, \binits{P.J.}},
\bauthor{\bsnm{Still}, \binits{T.}},
\bauthor{\bsnm{Lohr}, \binits{M.A.}},
\bauthor{\bsnm{Yodh}, \binits{A.}}:
\batitle{Suppression of the coffee-ring effect by shape-dependent capillary interactions}.
\bjtitle{nature}
\bvolume{476}(\bissue{7360}),
\bfpage{308}--\blpage{311}
(\byear{2011})
\end{barticle}
\endbibitem

\bibitem[\protect\citeauthoryear{Weber}{1855}]{weber1855mikroskopische}
\begin{barticle}
\bauthor{\bsnm{Weber}, \binits{E.H.}}:
\batitle{Mikroskopische beobachtungen sehr gesetzm{\"a}ssiger bewegungen, welche die bildung von niederschl{\"a}gen harziger k{\"o}rper aus weingeist begleiten}.
\bjtitle{Annalen der Physik}
\bvolume{170}(\bissue{3}),
\bfpage{447}--\blpage{459}
(\byear{1855})
\end{barticle}
\endbibitem

\bibitem[\protect\citeauthoryear{Scriven and Sternling}{1960}]{scriven1960marangoni}
\begin{barticle}
\bauthor{\bsnm{Scriven}, \binits{L.}},
\bauthor{\bsnm{Sternling}, \binits{C.}}:
\batitle{The marangoni effects}.
\bjtitle{Nature}
\bvolume{187}(\bissue{4733}),
\bfpage{186}--\blpage{188}
(\byear{1960})
\end{barticle}
\endbibitem

\bibitem[\protect\citeauthoryear{Xu and Luo}{2007}]{xu2007marangoni}
\begin{botherref}
\oauthor{\bsnm{Xu}, \binits{X.}},
\oauthor{\bsnm{Luo}, \binits{J.}}:
Marangoni flow in an evaporating water droplet.
Applied Physics Letters
\textbf{91}(12)
(2007)
\end{botherref}
\endbibitem

\bibitem[\protect\citeauthoryear{Hu and Larson}{2005}]{hu2005analysis}
\begin{barticle}
\bauthor{\bsnm{Hu}, \binits{H.}},
\bauthor{\bsnm{Larson}, \binits{R.G.}}:
\batitle{Analysis of the effects of marangoni stresses on the microflow in an evaporating sessile droplet}.
\bjtitle{Langmuir}
\bvolume{21}(\bissue{9}),
\bfpage{3972}--\blpage{3980}
(\byear{2005})
\end{barticle}
\endbibitem

\bibitem[\protect\citeauthoryear{Kim}{2022}]{kim2022multiple}
\begin{botherref}
\oauthor{\bsnm{Kim}, \binits{H.}}:
Multiple marangoni flows in a binary mixture sessile droplet.
Physics of Fluids
\textbf{34}(12)
(2022)
\end{botherref}
\endbibitem

\bibitem[\protect\citeauthoryear{Brutin}{2015}]{brutin2015droplet}
\begin{bbook}
\bauthor{\bsnm{Brutin}, \binits{D.}}:
\bbtitle{Droplet Wetting and Evaporation: from Pure to Complex Fluids}.
\bpublisher{Academic Press}, \blocation{???}
(\byear{2015})
\end{bbook}
\endbibitem

\bibitem[\protect\citeauthoryear{Thokchom and Shin}{2019}]{thokchom2019dynamical}
\begin{barticle}
\bauthor{\bsnm{Thokchom}, \binits{A.K.}},
\bauthor{\bsnm{Shin}, \binits{S.}}:
\batitle{Dynamical clustering and band formation of particles in a marangoni vortexing droplet}.
\bjtitle{Langmuir}
\bvolume{35}(\bissue{27}),
\bfpage{8977}--\blpage{8983}
(\byear{2019})
\end{barticle}
\endbibitem

\bibitem[\protect\citeauthoryear{Parsa et~al.}{2018}]{parsa2018mechanisms}
\begin{barticle}
\bauthor{\bsnm{Parsa}, \binits{M.}},
\bauthor{\bsnm{Harmand}, \binits{S.}},
\bauthor{\bsnm{Sefiane}, \binits{K.}}:
\batitle{Mechanisms of pattern formation from dried sessile drops}.
\bjtitle{Advances in colloid and interface science}
\bvolume{254},
\bfpage{22}--\blpage{47}
(\byear{2018})
\end{barticle}
\endbibitem

\bibitem[\protect\citeauthoryear{Wodlei et~al.}{2018}]{wodlei2018marangoni}
\begin{barticle}
\bauthor{\bsnm{Wodlei}, \binits{F.}},
\bauthor{\bsnm{Sebilleau}, \binits{J.}},
\bauthor{\bsnm{Magnaudet}, \binits{J.}},
\bauthor{\bsnm{Pimienta}, \binits{V.}}:
\batitle{Marangoni-driven flower-like patterning of an evaporating drop spreading on a liquid substrate}.
\bjtitle{Nature communications}
\bvolume{9}(\bissue{1}),
\bfpage{1}--\blpage{12}
(\byear{2018})
\end{barticle}
\endbibitem

\bibitem[\protect\citeauthoryear{Yamamoto et~al.}{2015}]{yamamoto2015evolution}
\begin{barticle}
\bauthor{\bsnm{Yamamoto}, \binits{D.}},
\bauthor{\bsnm{Nakajima}, \binits{C.}},
\bauthor{\bsnm{Shioi}, \binits{A.}},
\bauthor{\bsnm{Krafft}, \binits{M.P.}},
\bauthor{\bsnm{Yoshikawa}, \binits{K.}}:
\batitle{The evolution of spatial ordering of oil drops fast spreading on a water surface}.
\bjtitle{Nature communications}
\bvolume{6}(\bissue{1}),
\bfpage{1}--\blpage{6}
(\byear{2015})
\end{barticle}
\endbibitem

\bibitem[\protect\citeauthoryear{Chan and Fried}{2024}]{chan2024marangoni}
\begin{barticle}
\bauthor{\bsnm{Chan}, \binits{S.T.}},
\bauthor{\bsnm{Fried}, \binits{E.}}:
\batitle{Marangoni spreading on liquid substrates in new media art}.
\bjtitle{PNAS nexus}
\bvolume{3}(\bissue{2}),
\bfpage{059}
(\byear{2024})
\end{barticle}
\endbibitem

\bibitem[\protect\citeauthoryear{Baumgartner et~al.}{2022}]{baumgartner2022marangoni}
\begin{barticle}
\bauthor{\bsnm{Baumgartner}, \binits{D.A.}},
\bauthor{\bsnm{Shiri}, \binits{S.}},
\bauthor{\bsnm{Sinha}, \binits{S.}},
\bauthor{\bsnm{Karpitschka}, \binits{S.}},
\bauthor{\bsnm{Cira}, \binits{N.J.}}:
\batitle{Marangoni spreading and contracting three-component droplets on completely wetting surfaces}.
\bjtitle{Proceedings of the National Academy of Sciences}
\bvolume{119}(\bissue{19}),
\bfpage{2120432119}
(\byear{2022})
\end{barticle}
\endbibitem

\bibitem[\protect\citeauthoryear{Ma et~al.}{2023}]{ma2023experiments}
\begin{barticle}
\bauthor{\bsnm{Ma}, \binits{X.}},
\bauthor{\bsnm{Huang}, \binits{Y.}},
\bauthor{\bsnm{Huang}, \binits{Y.}},
\bauthor{\bsnm{Liu}, \binits{Z.}},
\bauthor{\bsnm{Li}, \binits{Z.}},
\bauthor{\bsnm{Floryan}, \binits{J.M.}}:
\batitle{Experiments on marangoni spreading--evidence of a new type of interfacial instability}.
\bjtitle{Journal of Fluid Mechanics}
\bvolume{958},
\bfpage{33}
(\byear{2023})
\end{barticle}
\endbibitem

\bibitem[\protect\citeauthoryear{Ma et~al.}{2020}]{ma2020fingering}
\begin{botherref}
\oauthor{\bsnm{Ma}, \binits{X.}},
\oauthor{\bsnm{Zhong}, \binits{M.}},
\oauthor{\bsnm{He}, \binits{Y.}},
\oauthor{\bsnm{Liu}, \binits{Z.}},
\oauthor{\bsnm{Li}, \binits{Z.}}:
Fingering instability in marangoni spreading on a deep layer of polymer solution.
Physics of Fluids
\textbf{32}(11)
(2020)
\end{botherref}
\endbibitem

\bibitem[\protect\citeauthoryear{Thayyil~Raju et~al.}{2022}]{thayyil2022evaporation}
\begin{barticle}
\bauthor{\bsnm{Thayyil~Raju}, \binits{L.}},
\bauthor{\bsnm{Diddens}, \binits{C.}},
\bauthor{\bsnm{Li}, \binits{Y.}},
\bauthor{\bsnm{Marin}, \binits{A.}},
\bauthor{\bsnm{Linden}, \binits{M.N.}},
\bauthor{\bsnm{Zhang}, \binits{X.}},
\bauthor{\bsnm{Lohse}, \binits{D.}}:
\batitle{Evaporation of a sessile colloidal water--glycerol droplet: Marangoni ring formation}.
\bjtitle{Langmuir}
\bvolume{38}(\bissue{39}),
\bfpage{12082}--\blpage{12094}
(\byear{2022})
\end{barticle}
\endbibitem

\bibitem[\protect\citeauthoryear{Bi et~al.}{2012}]{bi2012unconventional}
\begin{barticle}
\bauthor{\bsnm{Bi}, \binits{W.}},
\bauthor{\bsnm{Wu}, \binits{X.}},
\bauthor{\bsnm{Yeow}, \binits{E.K.}}:
\batitle{Unconventional multiple ring structure formation from evaporation-induced self-assembly of polymers}.
\bjtitle{Langmuir}
\bvolume{28}(\bissue{30}),
\bfpage{11056}--\blpage{11063}
(\byear{2012})
\end{barticle}
\endbibitem

\bibitem[\protect\citeauthoryear{Seo et~al.}{2017}]{seo2017altering}
\begin{barticle}
\bauthor{\bsnm{Seo}, \binits{C.}},
\bauthor{\bsnm{Jang}, \binits{D.}},
\bauthor{\bsnm{Chae}, \binits{J.}},
\bauthor{\bsnm{Shin}, \binits{S.}}:
\batitle{Altering the coffee-ring effect by adding a surfactant-like viscous polymer solution}.
\bjtitle{Scientific reports}
\bvolume{7}(\bissue{1}),
\bfpage{500}
(\byear{2017})
\end{barticle}
\endbibitem

\bibitem[\protect\citeauthoryear{Bennacer and Sefiane}{2014}]{bennacer2014vortices}
\begin{barticle}
\bauthor{\bsnm{Bennacer}, \binits{R.}},
\bauthor{\bsnm{Sefiane}, \binits{K.}}:
\batitle{Vortices, dissipation and flow transition in volatile binary drops}.
\bjtitle{Journal of fluid mechanics}
\bvolume{749},
\bfpage{649}--\blpage{665}
(\byear{2014})
\end{barticle}
\endbibitem

\bibitem[\protect\citeauthoryear{Kim et~al.}{2017}]{kim2017solutal}
\begin{barticle}
\bauthor{\bsnm{Kim}, \binits{H.}},
\bauthor{\bsnm{Muller}, \binits{K.}},
\bauthor{\bsnm{Shardt}, \binits{O.}},
\bauthor{\bsnm{Afkhami}, \binits{S.}},
\bauthor{\bsnm{Stone}, \binits{H.A.}}:
\batitle{Solutal marangoni flows of miscible liquids drive transport without surface contamination}.
\bjtitle{Nature Physics}
\bvolume{13}(\bissue{11}),
\bfpage{1105}--\blpage{1110}
(\byear{2017})
\end{barticle}
\endbibitem

\bibitem[\protect\citeauthoryear{Pyeon et~al.}{2022}]{pyeon2022self}
\begin{barticle}
\bauthor{\bsnm{Pyeon}, \binits{J.}},
\bauthor{\bsnm{Song}, \binits{K.M.}},
\bauthor{\bsnm{Jung}, \binits{Y.S.}},
\bauthor{\bsnm{Kim}, \binits{H.}}:
\batitle{Self-induced solutal marangoni flows realize coffee-ring-less quantum dot microarrays with extensive geometric tunability and scalability}.
\bjtitle{Advanced Science}
\bvolume{9}(\bissue{11}),
\bfpage{2104519}
(\byear{2022})
\end{barticle}
\endbibitem

\bibitem[\protect\citeauthoryear{Diddens et~al.}{2024}]{diddens2024non}
\begin{botherref}
\oauthor{\bsnm{Diddens}, \binits{C.}},
\oauthor{\bsnm{Dekker}, \binits{P.J.}},
\oauthor{\bsnm{Lohse}, \binits{D.}}:
Non-monotonic surface tension leads to spontaneous symmetry breaking in a binary evaporating drop.
arXiv preprint arXiv:2402.17452
(2024)
\end{botherref}
\endbibitem

\bibitem[\protect\citeauthoryear{Park et~al.}{2020}]{park2020control}
\begin{barticle}
\bauthor{\bsnm{Park}, \binits{J.}},
\bauthor{\bsnm{Ryu}, \binits{J.}},
\bauthor{\bsnm{Sung}, \binits{H.J.}},
\bauthor{\bsnm{Kim}, \binits{H.}}:
\batitle{Control of solutal marangoni-driven vortical flows and enhancement of mixing efficiency}.
\bjtitle{Journal of colloid and interface science}
\bvolume{561},
\bfpage{408}--\blpage{415}
(\byear{2020})
\end{barticle}
\endbibitem

\bibitem[\protect\citeauthoryear{Ryu et~al.}{2021}]{ryu2021analysis}
\begin{barticle}
\bauthor{\bsnm{Ryu}, \binits{J.}},
\bauthor{\bsnm{Kim}, \binits{J.}},
\bauthor{\bsnm{Park}, \binits{J.}},
\bauthor{\bsnm{Kim}, \binits{H.}}:
\batitle{Analysis of vapor-driven solutal marangoni flows inside a sessile droplet}.
\bjtitle{International Journal of Heat and Mass Transfer}
\bvolume{164},
\bfpage{120499}
(\byear{2021})
\end{barticle}
\endbibitem

\bibitem[\protect\citeauthoryear{Li et~al.}{1994}]{li1994sol}
\begin{barticle}
\bauthor{\bsnm{Li}, \binits{J.}},
\bauthor{\bsnm{Harada}, \binits{A.}},
\bauthor{\bsnm{Kamachi}, \binits{M.}}:
\batitle{Sol--gel transition during inclusion complex formation between $\alpha$-cyclodextrin and high molecular weight poly (ethylene glycol) s in aqueous solution}.
\bjtitle{Polymer journal}
\bvolume{26}(\bissue{9}),
\bfpage{1019}--\blpage{1026}
(\byear{1994})
\end{barticle}
\endbibitem

\bibitem[\protect\citeauthoryear{Baldwin and Fairhurst}{2014}]{baldwin2014effects}
\begin{barticle}
\bauthor{\bsnm{Baldwin}, \binits{K.}},
\bauthor{\bsnm{Fairhurst}, \binits{D.}}:
\batitle{The effects of molecular weight, evaporation rate and polymer concentration on pillar formation in drying poly (ethylene oxide) droplets}.
\bjtitle{Colloids and Surfaces A: Physicochemical and Engineering Aspects}
\bvolume{441},
\bfpage{867}--\blpage{871}
(\byear{2014})
\end{barticle}
\endbibitem

\bibitem[\protect\citeauthoryear{Asenjo and Andrews}{2012}]{asenjo2012aqueous}
\begin{barticle}
\bauthor{\bsnm{Asenjo}, \binits{J.A.}},
\bauthor{\bsnm{Andrews}, \binits{B.A.}}:
\batitle{Aqueous two-phase systems for protein separation: phase separation and applications}.
\bjtitle{Journal of Chromatography A}
\bvolume{1238},
\bfpage{1}--\blpage{10}
(\byear{2012})
\end{barticle}
\endbibitem

\bibitem[\protect\citeauthoryear{Grover and Ryall}{2005}]{grover2005critical}
\begin{barticle}
\bauthor{\bsnm{Grover}, \binits{P.K.}},
\bauthor{\bsnm{Ryall}, \binits{R.L.}}:
\batitle{Critical appraisal of salting-out and its implications for chemical and biological sciences}.
\bjtitle{Chemical reviews}
\bvolume{105}(\bissue{1}),
\bfpage{1}--\blpage{10}
(\byear{2005})
\end{barticle}
\endbibitem

\bibitem[\protect\citeauthoryear{Arakawa and Timasheff}{1984}]{arakawa1984mechanism}
\begin{barticle}
\bauthor{\bsnm{Arakawa}, \binits{T.}},
\bauthor{\bsnm{Timasheff}, \binits{S.N.}}:
\batitle{Mechanism of protein salting in and salting out by divalent cation salts: balance between hydration and salt binding}.
\bjtitle{Biochemistry}
\bvolume{23}(\bissue{25}),
\bfpage{5912}--\blpage{5923}
(\byear{1984})
\end{barticle}
\endbibitem

\bibitem[\protect\citeauthoryear{Hey et~al.}{2005}]{hey2005salting}
\begin{barticle}
\bauthor{\bsnm{Hey}, \binits{M.J.}},
\bauthor{\bsnm{Jackson}, \binits{D.P.}},
\bauthor{\bsnm{Yan}, \binits{H.}}:
\batitle{The salting-out effect and phase separation in aqueous solutions of electrolytes and poly (ethylene glycol)}.
\bjtitle{Polymer}
\bvolume{46}(\bissue{8}),
\bfpage{2567}--\blpage{2572}
(\byear{2005})
\end{barticle}
\endbibitem

\bibitem[\protect\citeauthoryear{Wu et~al.}{1996}]{wu1996interfacial}
\begin{barticle}
\bauthor{\bsnm{Wu}, \binits{Y.-T.}},
\bauthor{\bsnm{Zhu}, \binits{Z.-Q.}},
\bauthor{\bsnm{Mei}, \binits{L.-H.}}:
\batitle{Interfacial tension of poly (ethylene glycol)+ salt+ water systems}.
\bjtitle{Journal of Chemical \& Engineering Data}
\bvolume{41}(\bissue{5}),
\bfpage{1032}--\blpage{1035}
(\byear{1996})
\end{barticle}
\endbibitem

\bibitem[\protect\citeauthoryear{Qazi et~al.}{2017}]{qazi2017influence}
\begin{barticle}
\bauthor{\bsnm{Qazi}, \binits{M.J.}},
\bauthor{\bsnm{Liefferink}, \binits{R.W.}},
\bauthor{\bsnm{Schlegel}, \binits{S.J.}},
\bauthor{\bsnm{Backus}, \binits{E.H.}},
\bauthor{\bsnm{Bonn}, \binits{D.}},
\bauthor{\bsnm{Shahidzadeh}, \binits{N.}}:
\batitle{Influence of surfactants on sodium chloride crystallization in confinement}.
\bjtitle{Langmuir}
\bvolume{33}(\bissue{17}),
\bfpage{4260}--\blpage{4268}
(\byear{2017})
\end{barticle}
\endbibitem

\bibitem[\protect\citeauthoryear{Nasirimarekani et~al.}{2023}]{nasirimarekani2023pattern}
\begin{botherref}
\oauthor{\bsnm{Nasirimarekani}, \binits{V.}},
\oauthor{\bsnm{Ram{\`\i}rez-Soto}, \binits{O.}},
\oauthor{\bsnm{Karpitschka}, \binits{S.}},
\oauthor{\bsnm{Guido}, \binits{I.}}:
Pattern formation under mechanical stress in active biological networks confined inside evaporating droplets.
arXiv preprint arXiv:2305.07099
(2023)
\end{botherref}
\endbibitem

\bibitem[\protect\citeauthoryear{Falahati and Haji-Akbari}{2019}]{falahati2019thermodynamically}
\begin{barticle}
\bauthor{\bsnm{Falahati}, \binits{H.}},
\bauthor{\bsnm{Haji-Akbari}, \binits{A.}}:
\batitle{Thermodynamically driven assemblies and liquid--liquid phase separations in biology}.
\bjtitle{Soft matter}
\bvolume{15}(\bissue{6}),
\bfpage{1135}--\blpage{1154}
(\byear{2019})
\end{barticle}
\endbibitem

\bibitem[\protect\citeauthoryear{Shrinivas and Brenner}{2021}]{shrinivas2021phase}
\begin{barticle}
\bauthor{\bsnm{Shrinivas}, \binits{K.}},
\bauthor{\bsnm{Brenner}, \binits{M.P.}}:
\batitle{Phase separation in fluids with many interacting components}.
\bjtitle{Proceedings of the National Academy of Sciences}
\bvolume{118}(\bissue{45}),
\bfpage{2108551118}
(\byear{2021})
\end{barticle}
\endbibitem

\bibitem[\protect\citeauthoryear{Steiner et~al.}{1995}]{steiner1995entropy}
\begin{barticle}
\bauthor{\bsnm{Steiner}, \binits{U.}},
\bauthor{\bsnm{Meller}, \binits{A.}},
\bauthor{\bsnm{Stavans}, \binits{J.}}:
\batitle{Entropy driven phase separation in binary emulsions}.
\bjtitle{Physical review letters}
\bvolume{74}(\bissue{23}),
\bfpage{4750}
(\byear{1995})
\end{barticle}
\endbibitem

\bibitem[\protect\citeauthoryear{Zeuthen et~al.}{2019}]{zeuthen2019nanoparticle}
\begin{barticle}
\bauthor{\bsnm{Zeuthen}, \binits{C.M.}},
\bauthor{\bsnm{Shahrokhtash}, \binits{A.}},
\bauthor{\bsnm{Sutherland}, \binits{D.S.}}:
\batitle{Nanoparticle adsorption on antifouling polymer brushes}.
\bjtitle{Langmuir}
\bvolume{35}(\bissue{46}),
\bfpage{14879}--\blpage{14889}
(\byear{2019})
\end{barticle}
\endbibitem

\bibitem[\protect\citeauthoryear{Higaki et~al.}{2016}]{higaki2016anti}
\begin{barticle}
\bauthor{\bsnm{Higaki}, \binits{Y.}},
\bauthor{\bsnm{Kobayashi}, \binits{M.}},
\bauthor{\bsnm{Murakami}, \binits{D.}},
\bauthor{\bsnm{Takahara}, \binits{A.}}:
\batitle{Anti-fouling behavior of polymer brush immobilized surfaces}.
\bjtitle{Polymer Journal}
\bvolume{48}(\bissue{4}),
\bfpage{325}--\blpage{331}
(\byear{2016})
\end{barticle}
\endbibitem

\bibitem[\protect\citeauthoryear{Kreer}{2016}]{kreer2016polymer}
\begin{barticle}
\bauthor{\bsnm{Kreer}, \binits{T.}}:
\batitle{Polymer-brush lubrication: a review of recent theoretical advances}.
\bjtitle{Soft Matter}
\bvolume{12}(\bissue{15}),
\bfpage{3479}--\blpage{3501}
(\byear{2016})
\end{barticle}
\endbibitem

\bibitem[\protect\citeauthoryear{Giasson and Spencer}{2014}]{giasson2014aqueous}
\begin{botherref}
\oauthor{\bsnm{Giasson}, \binits{S.}},
\oauthor{\bsnm{Spencer}, \binits{N.D.}}:
Aqueous lubrication with polymer brushes.
Aqueous Lubrication,
183--218
(2014)
\end{botherref}
\endbibitem

\bibitem[\protect\citeauthoryear{Marin et~al.}{2019}]{marin2019solutal}
\begin{barticle}
\bauthor{\bsnm{Marin}, \binits{A.}},
\bauthor{\bsnm{Karpitschka}, \binits{S.}},
\bauthor{\bsnm{Noguera-Mar{\'\i}n}, \binits{D.}},
\bauthor{\bsnm{Cabrerizo-V{\'\i}lchez}, \binits{M.A.}},
\bauthor{\bsnm{Rossi}, \binits{M.}},
\bauthor{\bsnm{K{\"a}hler}, \binits{C.J.}},
\bauthor{\bsnm{Rodr{\'\i}guez~Valverde}, \binits{M.A.}}:
\batitle{Solutal marangoni flow as the cause of ring stains from drying salty colloidal drops}.
\bjtitle{Physical review fluids}
\bvolume{4}(\bissue{4}),
\bfpage{041601}
(\byear{2019})
\end{barticle}
\endbibitem

\end{thebibliography}
\newpage
\section*{Supplementary Figures}
\setcounter{figure}{0} 
\renewcommand{\figurename}{Figure S.}
\subsection*{Effect of droplet size on the final pattern formation by phase separation bursting flow}
The flower-like pattern (the droplet with an initial PEG concentration of 1\% w/w) was observed as a random observation that sometimes occurs and shows some degree of change in different experiments and does not represent a reproducible pattern. We believe that a non-homogeneous surface condition (a degree of defects in the polymer brushes) and localized impurities when conducting the experiments in the room conditions lead to a partial disruption of the vortices and the observation of a complete flower-like pattern. Furthermore, we asked the question whether the droplet size has an influence on the formation of a complete flower-like pattern. The results show that for smaller droplets the petals are more compact, but the number of vortices seems to remain similar (Figure S1.a). Furthermore, the comparison of two different droplet sizes for the 3\% PEG droplet shows that the droplet size has no effect on the final pattern formation when the phase separation bursting flow shows random deposition of the partciles (Figure S1.b). From these results, we conclude that the spatial freedom in the formation of the flower-like pattern due to the vortices plays an important role in the patterning of the beads during phase separation bursting flow. 

\begin{figure*}[h!]
    \centering
    \includegraphics[width=0.8\textwidth]{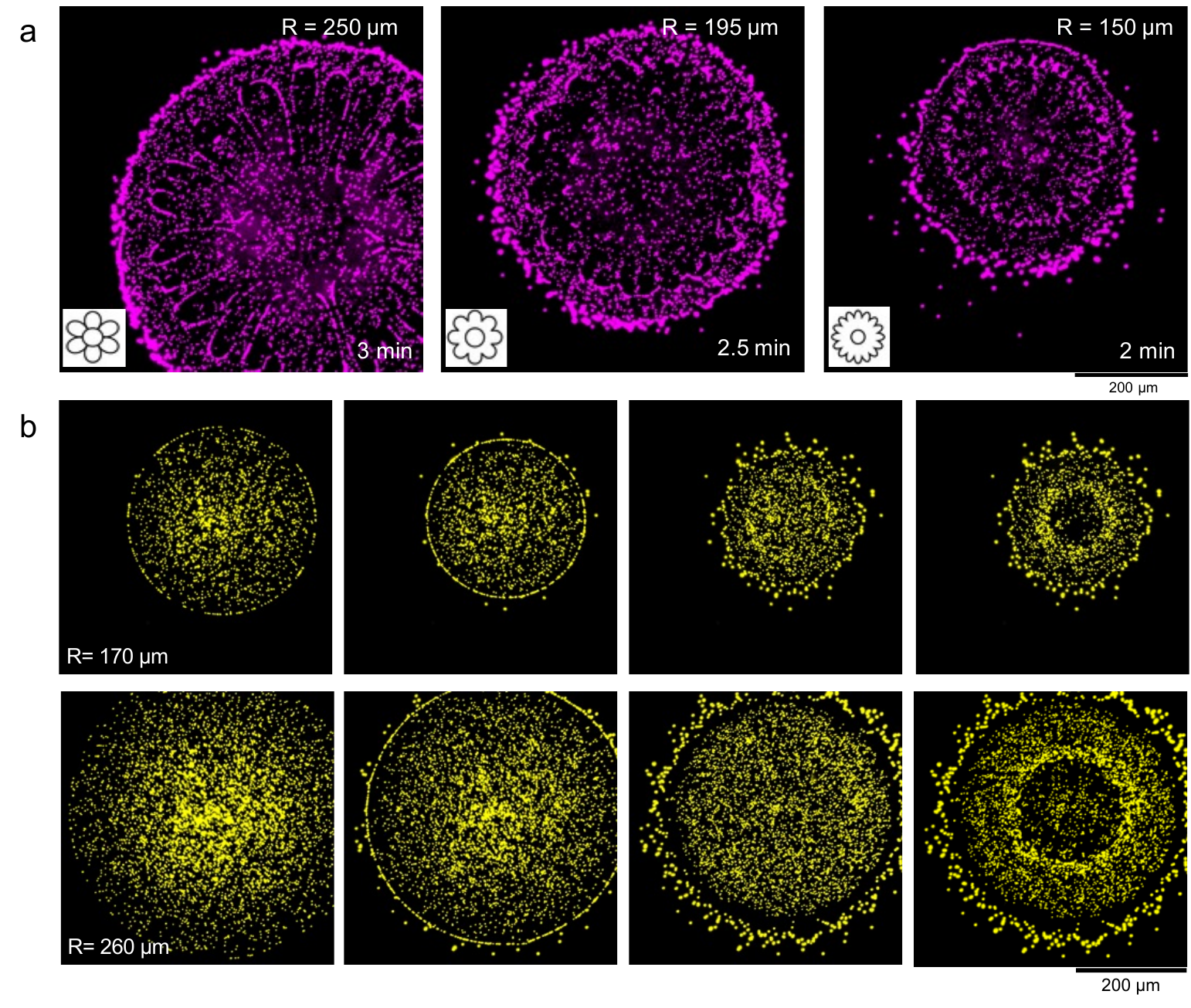} 
    \caption{Effect of droplet size on the final pattern inside the droplet showing phase separation bursting flow. a) Microscopy images of the three different droplets with different initial sizes showing some change in the final flower-like pattern. b) Comparison of the final pattern in two droplets of two different initial sizes with an initial PEG concentration of 3\% w/w.}
    \label{fig:figS2}
\end{figure*}
\subsection*{Investigation of the formation of solutal vortices in a salt-containing droplet using quantum dots as tracer particles}
The question was asked whether vortices could be observed in the salt-containing droplet in the absence of phase separation of the polymeric monomers within the droplet. First, we mixed M2B buffer with 2 $ \mu m$ beads and subjected the droplet to evaporation, which showed no vortices throughout the drying process. Considering that the flow generated by the solutal Marangoni flows may not be strong enough to move the micrometer-sized beads, we also performed the same experiment with quantum dots with an average size of 15 nm (Figure S2). The experiment with quantum dots also shows no vortex formation, and a similar deposition of particles is observed in the salt-enriched region (II) of the droplet. Although certain have been shown to form vortices~\cite{kim2017solutal,marin2019solutal}, the buffer used here, shows no visible vortices. 
\begin{figure*}[h!]
    \centering
    \includegraphics[width=0.6\textwidth]{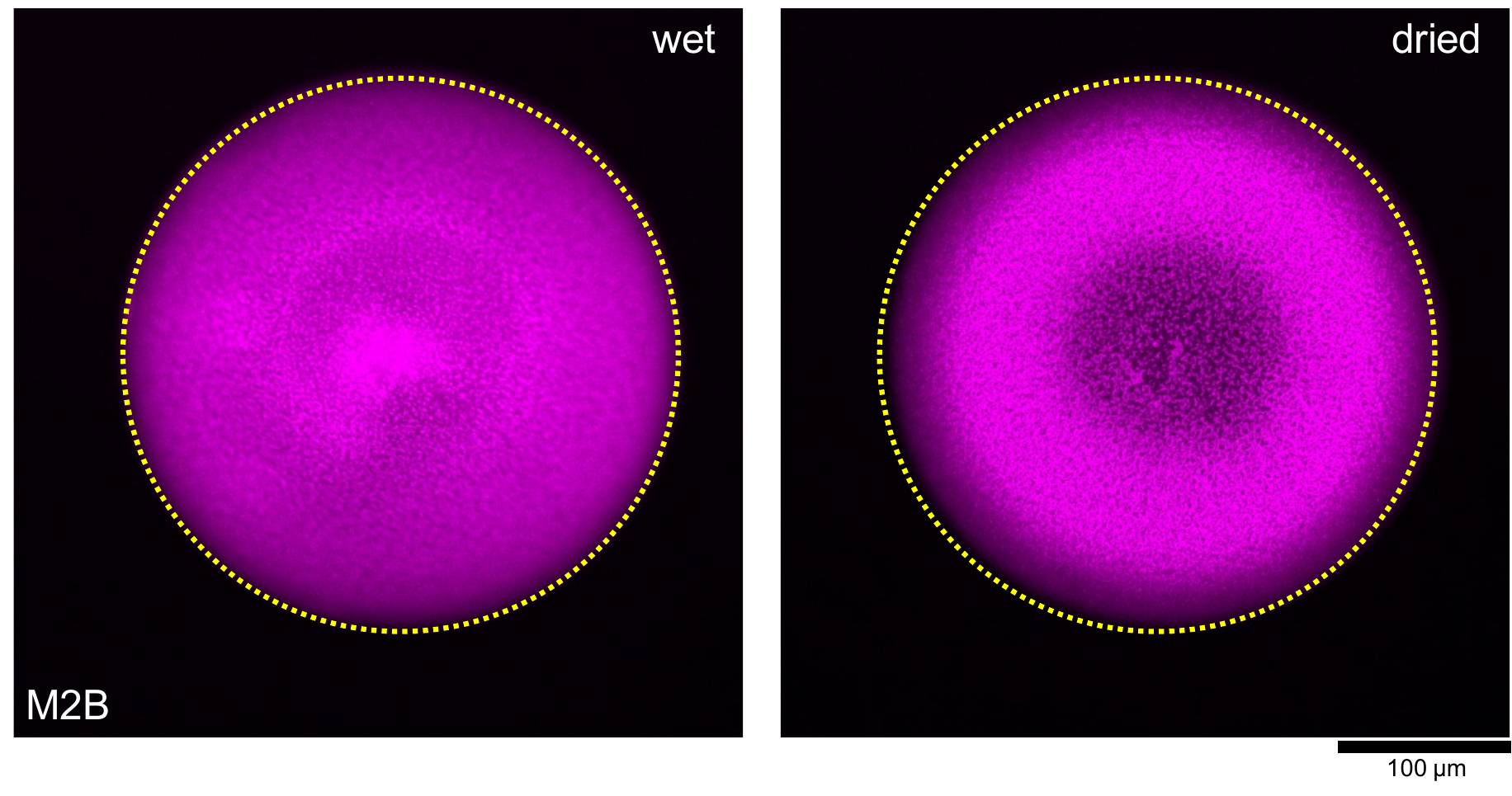} 
    \caption{Deposition pattern of the quantum dots during the evaporation of the salt-containing M2B buffer. The yellow dashed line represents the initial contact line of the droplet. }
    \label{fig:fig1}
\end{figure*}
\newpage
\subsection*{Studying the effects of the substrate on the Marangoni and phase separation flows inside the droplet}
It was observed that the phase separation leads to a fan-shaped deposition of the particles in the center of the droplet when the droplet is deposited on a clean glass substrate instead of a glass substrate functionalized with polymer brushes (Figure S3.a,b). In addition, coffee ring deposition of the particles was also observed due to the interaction of the particles with the substrate (Figure S3.c). We have asked whether the interactions of the particles with the substrate change the two main flows, Marangoni and phase separation bursting. Particle tracking within a droplet with an initial PEG concentration of 1\% w/w shows that both flows are active and clearly visible inside the droplet throughout the drying process. We conclude that although the interaction of the beads with the glass substrate changes the final pattern, the formation of flows is not influenced by the surface properties of the substrate. More importantly, phase separation bursting flow occurs independently of the surface properties of the substrate. 
\begin{figure*}[h!]
    \centering
    \includegraphics[width=\textwidth]{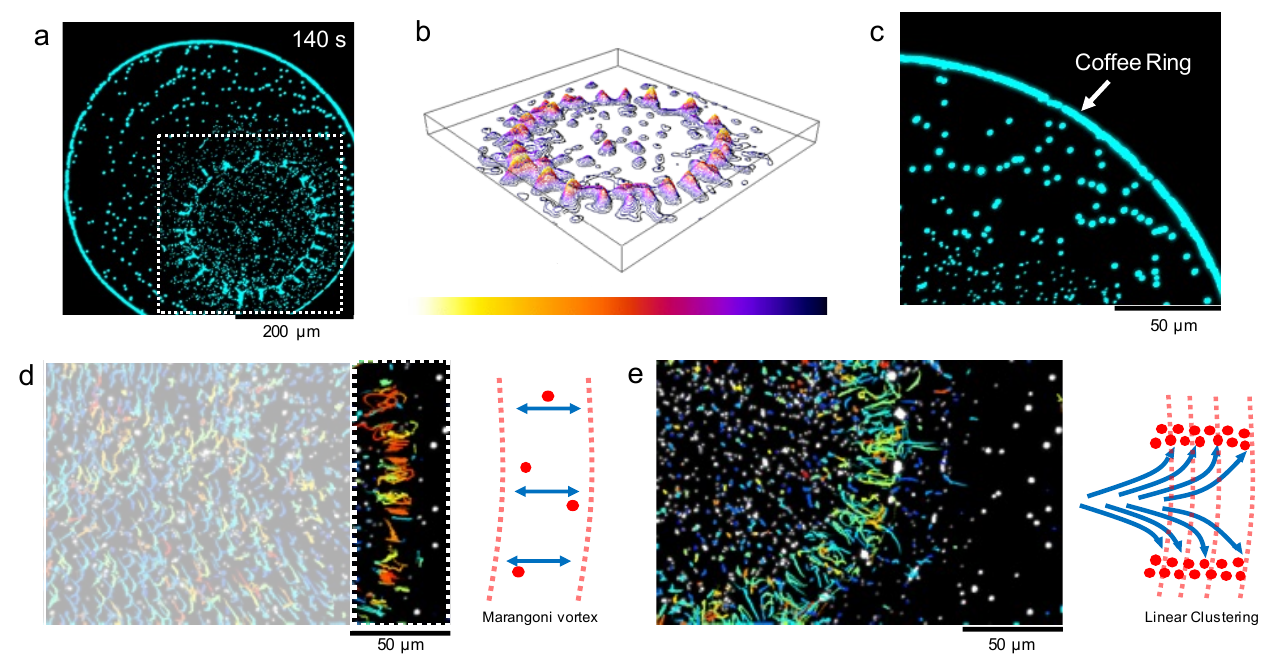} 
    \caption{Effect of substrate interactions on the formation of the final pattern of a 1\% w/w PEG droplet. a) Microscopy image of the droplet forming a fan-shaped deposition of the particles in the center of the droplet as a result of the phase separation bursting flow. b) 3D plot based on the light intensity of the deposited particles shows that the particles are clustered. c) Inset view highlighting the formation of the coffee ring. d) shows the particle tracking data indicating the occurrence of Marangoni flow in the early stages of vaporization. The schematic illustrates that the particles move back and forth in a vortex. e) Observation of phase separation bursting flow in later stages of evaporation.}
    \label{fig:fig3}
\end{figure*}



\end{document}